\documentclass[11pt]{JHEP3} 

\usepackage{amsmath}
\usepackage{amsfonts}

\usepackage{epsfig}
\usepackage{amssymb}

\newcommand{\beq}{\begin{eqnarray}}
\newcommand{\eeq}{\end{eqnarray}}
\newcommand{\be}{\begin{eqnarray}}
\newcommand{\ee}{\end{eqnarray}}

\newcommand{\LCDM}{$\Lambda$CDM }

\newcommand{\strong}[1]{\textbf{#1}}

\title{Disformal scalar fields and the dark sector of the universe}

\author{M. Zumalac\'arregui \\ 
Institut de Ciencies del Cosmos, Universitat de Barcelona \\ Marti i Franques 1, E-08028 Barcelona, Spain \\
E-mail: \email{miguelzuma@am.ub.es}}
\author{T. S. Koivisto \\ 
Institute for Theoretical Physics and the Spinoza Institute, Utrecht University, \\
Leuvenlaan 4, Postbus 80.195, 3508 TD Utrecht, The Netherlands \\
E-mail: \email{t.s.koivisto@uu.nl}}
\author{D. F. Mota \\ 
Institute of Theoretical Astrophysics, University of Oslo, 0315 Oslo, Norway \\
E-mail: \email{d.f.mota@astro.uio.no}}
\author{P. Ruiz-Lapuente \\ 
Institut de Ciencies del Cosmos, Universitat de Barcelona \\ Marti i Franques 1, E-08028 Barcelona, Spain \\
E-mail: \email{pilar@am.ub.es}}

\abstract{Disformal transformations have proven to be very useful to devise models of the dark sector.
In the present paper we apply such transformation to a single scalar field theory as a way to drive the field into a slow roll phase. 
The canonical scalar field Lagrangian, when coupled to a disformal metric, turns out to have relations to bimetric dark matter theories and 
to describe many specific dark energy models at various limits, thus providing a surprisingly simple parametrisation of a wide variety of 
models including tachyon, Chaplygin gas, K-essence and dilatonic ghost condensate.
We investigate the evolution of the background and linear perturbations in disformal quintessence in order to perform a full comparison of 
the predictions with the cosmological data. 
The dynamics of the expansion, in particular the mechanism of the transition 
to accelerating phase, is described in detail. We then study the effects of disformal quintessence on cosmic microwave background (CMB) 
anisotropies and large scale structures (LSS). A likelihood analysis using the latest data on wide-ranging SNIa, CMB and LSS 
observations is performed allowing variations in six cosmological parameters and the two parameters specifying the model.
We find that while a large region of parameter space remains compatible with observations, models featuring either too much early dark 
energy or too slow transition to acceleration are ruled out.  
}

\begin{document}

\section{Introduction}

Observations of the Cosmic Microwave Background temperature anisotropy reveal that
a mysterious constituent with negative pressure, so called dark energy, accounts for about $70\%$
percent of today's mass energy budget and is causing the expansion of the universe to
accelerate \cite{wmap,lss}. These observations are in remarkable concordance with the observations of
distant supernovae \cite{sn,sn1}. A major challenge in present day cosmology is to discover the physical nature of
dark energy. Meanwhile, though the evidence for the existence of dark matter has been accumulating for several 
decades, the quest to find its precise nature, whether as a particle described by an extension of the standard model, 
a more exotic field or even a modification of General Relativity, is ongoing. Many possibilities have been explored
in attempts to explain the workings of the dark sector \cite{copeland,Durrer:2007re,koi1,koi2,Li1,koi3,aether,koi4,Li2,koi5,li,lli1}.


The phenomenologically simplest but theoretically very problematical dark energy is the Einsteins $\Lambda$ term \cite{Nobbenhuis:2006yf}.
Quintessence provides a dynamical alternative to the static cosmological constant \cite{Wetterich:1987fm,Peebles:1987ek}. 
It tracks or scales with the background energy density, therefore more naturally resulting in similar orders of magnitude for
the dark energy and dust-like matter energy densities today. However, one needs some mechanism to end this scaling
and to onset the acceleration. Several possibilities have been considered, in particular
introducing a suitable bump into the form potential \cite{Albrecht:1999rm},
taking into account the Gauss-Bonnet term \cite{Koivisto:2006ai,Koivisto:2006xf},
coupling the field to other matter \cite{Amendola:1999er,Koivisto:2005nr},
considering non-canonical Lagrangian \cite{ArmendarizPicon:2000ah},
or introducing many fields \cite{Copeland:2000vh}. In the following, a disformal relation \cite{Bekenstein:1992pj} is applied for this 
purpose. 

Disformal relations have recurrently appeared in cosmology in models of 
alternative dark matter, in particular TeVes \cite{teves} and other relativistic and covariant formulations of MOND \cite{mond}, and in 
bigravity theories \cite{Banados:2008rm}. This comes about for several reasons: theoretically, it is related to the Born-Infeld type of 
Lagrangians (as will become clear below), and phenomenologically, it is needed to produce the observed lensing without resorting to
particle dark matter \cite{Skordis:2009bf}. Let us mention that the possible relevance to inflation has also been considered 
\cite{Kaloper:2003yf}, and that disformality is a key to varying speed of light, which has been considered as an alternative to
inflation \cite{Clayton:1998hv,Clayton:2001rt}. The usefulness of conformal transformations is well
known and appreciated \cite{Faraoni:2006fx,Kastrup:2008jn,Dabrowski:2008kx}, but many aspects of the wider framework remain 
uncovered. The conformal transformation relates the simplest scalar-tensor theories (including the $f(R)$ theories 
\cite{DeFelice:2010aj}) to general relativity \cite{Magnano:1993bd}, but in the general case one is prompted to turn into extended 
classes of transformations. 

Here we make a preliminary step along this direction by exploring the perhaps simplest possible set-up: the canonical scalar field 
coupled to a disformal metric. It turns that already then many relations between seemingly disconnected models emerge. If the scalar field 
is reduced to a constant term, it appears as a tachyon or a Chaplygin gas \cite{Kamenshchik:2001cp} in the physical frame. When the 
kinetic term is included, some other previously considered models can be recovered, as we will discuss in Section \ref{dm}. In the present 
study we then focus on one simple 
possibility, disformal quintessence. We point out that it can act as viable dark energy when the parameters in the Lagrangian are of the 
Planck scale. Section \ref{dq} is devoted to study of this model, and contains the main results of this paper. 
We describe in detail the background evolution of this model in subsection \ref{background-section}. Particular attention is given to the 
transition mechanism providing an exit from the scaling era. The details of this transition depend on the two parameters of the model, and 
thus they can be constrained by the SNeIa data. We also consider the evolution of perturbations in subsection \ref{perturb-section} in 
order to compute the CMB and matter power spectra. Armed with these solutions, we perform a Monte Carlo analysis of the model in subsection 
\ref{constraints-section}, combining data from many different cosmological experiments. Wide parameter range is found to be compatible with 
observations, but certain parameter region, corresponding to 
shallow slope of the exponential potential or of the disformal factor, can be ruled out. We conclude briefly in section 
\ref{conclusions-section}.

\section{Dark energy and unified models from disformal relation}
\label{dm}

A generic scalar field Lagrangian is a function of two variables,
\be
\mathcal{L} = \mathcal{L}(\phi,X). 
\ee
Expanded as a series in the kinetic term $X=\frac{1}{2}g^{\mu\nu}\phi_{,\mu}\phi_{,\nu}$ it reads
\be \label{expand}
\mathcal{L} = \mathcal{L}(\phi,0) + \mathcal{L}_{,X}(\phi,0)X + \mathcal{L}_{,XX}(\phi,0)X^2 + \dots
\ee
The first term corresponds to the potential, and the second term, if canonically normalized, equals one.
The dimension of the prefactor in third term is $1/M^2$, and the relative importance of this term seems to be
$H/M$. Unless the the prefactor is huge, this term is negligible in the late universe. Apparently a canonical scalar field should be an 
adequate description of almost any natural case. However, we will soon see that this can be avoided for suitable forms of $\mathcal{L}$.
This comes about simply because we can have an exponential amplification of the coefficients in the formal series (\ref{expand}). 

We consider the following generalization of the conformal transformation   
\be \label{disf}
\bar{g}_{\alpha\beta} = A(\phi)g_{\alpha\beta} + B(\phi) \phi_{,\alpha}\phi_{,\beta}
\ee
defined by a scalar field $\phi$. Equation (\ref{disf}) constitutes the most general relation between metrics that preserves causality and covariance and is defined through $\phi$ and its first derivatives only \cite{Bekenstein:1992pj}.

\subsection{Disformal dark energy} \label{chaplygin-section}

The simplest theory where (\ref{disf}) appears to be relevant is
\be\label{chaplygin}
S_{\bar{\Lambda}} = -\int d^n x \sqrt{-\bar{g}}\Lambda.
\ee
It turns out that such an action corresponds to a matter component obeying the equation of state
\be \label{chap}
p = - \frac{\Lambda^2}{\rho}.
\ee
Thus the cosmological term of a disformal metric is a tachyon in a constant potential\footnote{The action given by (\ref{chaplygin}) can be easily written in terms of the gravitational metric using (\ref{hatted}) for the determinant of the disformal metric.}. This describes unstable branes
in string theory \cite{Sen:2002nu,Sen:2002in}. The equation of
state (\ref{chap}) corresponds to Chaplygin gas , which has been applied in cosmology in attempts to unify dark matter
and dark energy \cite{Kamenshchik:2001cp}. Our result shows that this rather exotic fluid can be obtained in an extremely simple way from a
cosmological constant term by taking into account its  possible disformal relation to the matter metric.

The $\Lambda$ term describes the potential of the tachyon, and therefore it is natural to consider it as a field dependent term.
A straightforward generalization of this is to include also a kinetic
term for the field. In the case of canonical kinetic term 
we obtain the model
\be \label{sphi}
S_{\phi} = -\int d^n x \sqrt{-\bar{g}}\left(\frac{\varepsilon}{2}\bar{g}^{\mu\nu}\phi_{,\mu}\phi_{,\nu}+V(\phi)\right).
\ee
As it does not complicate our analysis we keep the sign of the kinetic term, $\varepsilon$ arbitrary.
The case $\varepsilon=1$ was considered in \cite{Koivisto:2008ak} as a variation of the quintessence scenario. The effect of the disformal relation 
is then to end the scaling era and begin the accelerated expansion. The canonical field in the disformal metric assumes the rather unappealing form in terms of the 
physical metric
\be
\mathcal{L}_\phi & = & A^\frac{3}{2}(\phi)\left(\frac{\varepsilon X}{\sqrt{A(\phi)+2B(\phi)X}} 
 +  \sqrt{A(\phi)+2B(\phi)X}V(\phi)\right). \label{unbarred}
\ee
For derivation of this form and a variation of the model, see subsection \ref{vari}. Then, expansion in the kinetic energy follows
\begin{eqnarray}\label{expand2}
\mathcal{L}_\phi &=& A^2V \, + \, AX (\varepsilon+BV) \, - \, \frac{1}{2}BX^2 (2\varepsilon+BV) \\  
&&\, + \,\frac{ B^2 X^3 }{2 A}  (3\varepsilon+BV) \, + \,\frac{ 5 B^3 X^4 }{4 A^2}  (2\varepsilon+BV)
\, + \,\cdots  \nonumber
\end{eqnarray}
Due to the form of the expansion all terms after first order in $X$ are of the form $X^n B^{n-1} A^{2-n} $ $ [p_n \varepsilon + q_n BV]$ with $p_n,q_n$ rational numbers. The argument following (\ref{expand}) can be circumvented if the coefficients are adequately enhanced so that the expansion is not valid anymore. One possibility is to choose an exponential form for the disformal function $B$ (with $A=1$).

Notice that the relation (\ref{disf}) becomes meaningful only after we have specified the coupling to gravity. We consider the gravity sector to be given by the Einstein-Hilbert term of the unbarred metric. Then action (\ref{sphi}) could be considered to be the Einstein frame version of a scalar tensor theory with derivative couplings \cite{luca}. In terms of the Jordan frame metric, the variation of the action (\ref{sphi}) alone gives the standard equation of motion of the field, $\bar{\Box}\phi + V'(\phi) = 0$, and the canonical stress tensor - however, one would have to take into account that the gravity piece would then be nonminimally coupled to the scalar field, resulting considerably more involved equations.

As we have seen, the disformal relation has a very simple connection with the Chaplygin gas. It turns out that the theory given by (\ref{sphi},\ref{unbarred}) can encompass a wide range of dark energy models through different choices of $\varepsilon$, $B$, $V$ and $A$. Some variations are shown in Table \ref{othermodels}. Note that the very inclusion of the disformal metric $\bar g_{\mu\nu}$ makes the scalar field dynamical, as it contains derivatives of $\phi$. Disformal quintessence will be discussed in depth in section \ref{disformal-section}.

\TABULAR{c  c  c  l  l}{
$\varepsilon$ & $B(\phi)$ & $V(\phi)$ &\ \ Model & \  Remarks \\ \hline
0 & 0 & $\Lambda$ & Cosmological constant & $w=-1$  \\
1 & 0 & $V(\phi)$ & Quintessence & $w>-1$, \cite{Wetterich:1987fm,Peebles:1987ek} \\
-1 & 0 & $V(\phi)$ & Phantom quintessence & $w<-1$, \cite{Caldwell:1999ew} \\
\hline
0 & $>0$ & $\pm\Lambda$ & (Anti-)Chaplygin gas & Eq. \ref{chaplygin}, \cite{Kamenshchik:2001cp} \\
0 & $1$ & $V(\phi)$ & Tachyon condensate & \cite{Padmanabhan:2002cp,Padmanabhan:2002sh} \\
-1 & $A(\phi)$ & 0 & K-essence \footnotemark[1] & $A=B>0$, \cite{Chiba:1999ka,ArmendarizPicon:2000ah} \\
-1 & $e^{\beta\phi}$ & 0 & Dilatonic ghost\footnotemark[1] & \cite{Piazza:2004df,ArkaniHamed:2003uz} \\ 
\hline
-1 & - & - & Disformal phantom & \\
1 & $>0$ & $V(\phi)$ & Disformal quintessence & \cite{Koivisto:2008ak}, Section \ref{disformal-section}  \\ 
\hline
\footnotetext[1]{Assuming that the expansion (\ref{expand2}) is consistent to second order in $X$.}
}{Disformal dark energy models obtained from the action (\ref{sphi}). 
The first three models do not use disformal transformation. The second set contains examples which have been thoroughly studied in the literature. Disformal quintessence will be considered further in this work, while the phantom version is left for further studies. All models are purely disformal ($A=1$) unless explicitly mentioned. A modification to obtain further variations is briefly discussed in subsection \ref{vari}.\label{othermodels}}



\subsection{Bigravity}

Recently the Born-Infeld form of the action has been considered in the framework of bigravity theories \cite{Banados:2008rm}.
Following Ref.\cite{Banados:2008fi}, let us write
\be \label{ebi1}
S  =  S_{EH}(g) + C S_{EH}(q)  +  \int\sqrt{-q}d^n x\left(\mathcal{L}_\phi-\frac{1}{2}A q^{\mu\nu}g_{\mu\nu}\right),
\ee
where the two first terms are the Einstein-Hilbert actions of the metric $g$ and the metric $q$, respectively, with a relative strength 
given by $C$.
The interaction term coupled to the metric $q$ includes an additional scalar field as
\be
\mathcal{L}_\phi = \frac{1}{2}Bq^{\mu\nu}\phi_{,\mu}\phi_{,\nu} + U(\phi).
\ee
In general, $A$, $B$ and $C$ could be functions of $\phi$, though we do not write this explicitly. By varying with respect to $q^{\mu\nu}$, 
one finds
the relation of the determinants
\be
q\left(\mathcal{L}_\phi  -  \frac{1}{2}A q^{\mu\nu}g_{\mu\nu}+\frac{1}{2}C K\right)^n = \tilde{g},
\ee
where $K_{\mu\nu}$ is the Ricci tensor and $K$ the Ricci scalar associated to $q_{\mu\nu}$ and $\tilde{g}$ is the determinant of the metric
\be
\tilde{g}_{\mu\nu} = Ag_{\mu\nu}+B\phi_{,\mu}\phi_{,\nu}-CK_{\mu\nu}.
\ee
Using the trace of the field equations for $q_{\mu\nu}$ we further get that
\be
q = \left(\frac{n-2}{2nU}\right)^n\tilde{g}.
\ee
This leads us to the observation that the scalar-bitensor theory (\ref{ebi1}) we started with is equivalent to
\be
S = S_{EH}(g) + \int d^n x \sqrt{-\tilde{g}} \left(\frac{n-2}{2nU}\right)^{\frac{n}{2}-1}.
\ee
Setting $A=U=1$, $B=0$ we get the Einstein-Born-Infeld (EBI) theory. This slightly generalizes the derivation
presented in \cite{Banados:2008fi}, where the equivalence of a bimetric theory and the EBI theory
was pointed out. It is interesting that the disformal scalar field enters here into play, since up to now the EBI
has been shown to mimic dark matter both at galactic and cosmological scales, but the structure formation having
tension with the data when simultaneously providing dark energy \cite{Banados:2008fj}. The Chaplygin gas ensuing
from the tachyon Lagrangian on the other hand can act as dark energy but the structure formation bounds rule it
out as a candidate for dark matter \cite{Sandvik:2002jz,Koivisto:2004ne}. This suggests that the successful
unification of the dark sector ultimately might require considering a theory along the lines of
(\ref{ebi1}). In the present study, we make a step towards this by analyzing in detail the scalar sector of such a
theory, which could be responsible for dark energy.

\subsection{Rapid transition}

We also note that a solution to the structure formation problem of the unified models discussed above has been suggested based
on a rapid transition of the unified fluid from a dark matter-like phase to an effective cosmological 
constant \cite{Piattella:2009kt}. Then, the 
sound speed could be nearly vanishing except for the transition period that is very short in cosmological time scales. As we will 
show in detail in section 
\ref{transition-section}, the disformal relation can naturally provide means for such a sharp transition, thus suggesting that a simple 
field theory may generate the desired scenarios. However, to assess their viability, one would have to consider the full nonlinear 
evolution of cosmological structures, since the linear perturbation theory predicts that when the cosmological constant phase 
is reached, the structures that have formed earlier are simply wiped away, resulting in vanishing matter power today. In reality though 
the already virialized objects may be expected to retain their form as the field changes its nature at the very largest cosmological scales, 
but the details of this are beyond the scope of our present study. 

\subsection{A variation of the model} \label{vari}
   
By contraction, the inverse disformal metric can be shown to be given by
\be
\bar{g}^{\alpha\beta}=\frac{1}{A}\left(g^{\alpha\beta}-\frac{B}{A+2BX}\phi^{,\alpha}\phi^{,\beta} \right)\,.
\ee
Then we see that
\be
\bar{X} = \frac{X}{A+2BX}\,.
\ee
The determinant of the disformal metric is (for a method to obtain the result see Appendix C of Ref.\cite{Bekenstein:2004ne})
\be
\sqrt{-\bar{g}} = A^2\sqrt{1+2\frac{B}{A}X}\sqrt{-g}\,. \label{hatted}
\ee
Using these formulas the form (\ref{unbarred}) follows immediately.


One may also consider an alternative prescription for the canonical field (\ref{sphi}). The kinetic term is written there is terms of the 
inverse metric 
as $\bar{X}=\frac{1}{2}\bar{g}^{\mu\nu}\phi_{,\mu}\phi_{,\nu}$. The alternative formulation employs the metric in the combination 
$\hat{X}=\frac{1}{2}\bar{g}_{\mu\nu}\phi^{,\mu}\phi^{,\nu}$. This is a consistent but not the minimal prescription, since depending on the 
viewpoint, we 
are doing one of the following: 1) mixing the two metrics (since the derivative indices in (\ref{hatted}) are raised with the unbarred 
metric) 2) 
considering non-canonical field (the unbarred metric can be barred by introducing field combinations) 3) defining the kinetic energy in terms 
of 
differentials with respect to one-forms (and not the coordinate vector as usually). Now $\hat{X}=X(1+2BX)$, and in terms of the matter 
metric, this 
Lagrangian corresponding to this model is
\be
\hat{\mathcal{L}}_\phi  =  A^\frac{3}{2}\sqrt{A+2BX}\left[X(1+2BX)+V(\phi)\right]\,. \label{hatted_l}
\ee
We shall not consider this variation further here.

\section{Disformal quintessence}\label{disformal-section}
\label{dq}

We will now explore the model given by action (\ref{sphi}) and (\ref{unbarred}) with $\varepsilon =1$. For the sake of simplicity, we will restrict to the purely disformal case $A=1$. The dynamical equations derived in \ref{background-section} and \ref{perturb-section} hold for any choice of $B$ and $V$, but we will rather focus on an exponential dependence for both functions
\begin{equation}\label{B}
B ={M_p^{-4}} \ e^{\beta(\phi+\phi_x)/M_p}\,,
\end{equation}
\begin{equation} \label{V}
V = M_p^4 \ e^{-\alpha\phi/M_p}\,.
\end{equation}

The model parameters are chosen such that the disformal features are negligible at early times so the field behaves as normal quintessence. 
The form for $V$ ensures the existence of scaling solutions in this regime if $\alpha>2$ \cite{Copeland:1997et}. 
The exponential form for $B$ allows the disformal features to become relevant without introducing a new scale. 


\subsection{Background evolution} \label{background-section}

For a flat FRW metric, the energy and density pressure in the Einstein frame read \cite{Koivisto:2008ak}
\begin{eqnarray}\label{rho}
\rho = \frac{1}{\sqrt{\mathcal L}}\Big(\varepsilon \frac{\dot\phi^2}{2\mathcal L}+ V \Big) \\ \label{pressure}
p = \sqrt{\mathcal L}\Big(\varepsilon \frac{\dot\phi^2}{2\mathcal L}- V \Big) 
\end{eqnarray}
where the \emph{lapse function} 
\begin{equation}
\mathcal L = 1-B\dot\phi^2 \, ,
\end{equation}
measures the relative time flow in the disformal frame (\ref{disf}) relative to the Einstein frame i.e. $\bar g_{00}/g_{00}$. 
The \emph{disformal factor}
\begin{equation}
\mathcal D = B\dot\phi^2 \, ,
\end{equation}
is a measurement of the deviation with respect to canonical quintessence.

The nonlinear field equation can be written in a form analogous to a single harmonic oscillator with coefficients that depend on  $\phi,\dot\phi$
\begin{equation}\label{equation}
\ddot\phi + 3 \frac{\mathcal F}{ \mathcal M} H \dot\phi + \frac{\mathcal P}{ \mathcal M} =0 \,.
\end{equation}
The analogs of mass, friction and potential terms can be written in terms of the disformal factor and the lapse as
\begin{eqnarray}
\mathcal M &=& \varepsilon (1 +\frac{1}{2}\mathcal D) +\mathcal L BV  \label{Mbckg}\,, \\
\mathcal F &=& \mathcal L [\varepsilon (1 -\frac{1}{2}\mathcal D) + \mathcal L BV ]\,, \label{Fbckg}  \\
\mathcal P &=&[\varepsilon \frac{3}{4}\dot\phi^2+\frac{1}{2}\mathcal L V] B_{,\phi}\dot\phi^2 +\mathcal L^2 V_{,\phi}\,.  \label{Pbckg} 
\end{eqnarray}
In addition to the disformal factor and the lapse, we find the dimensionless \emph{disformal-potential factor} $BV$ as yet another measurement of the disformal properties.
\begin{equation}\label{BV}
BV = e^{[(\beta-\alpha){\phi} + \beta{\phi_x}]/M_p} 
\end{equation}

\subsubsection{Tracking}\label{trackingstage}

We assume that the field has reached the tracking solution with $\Omega_\phi = n/\alpha^2$ before any relevant time. This behavior will last while $\mathcal D\ll1$ and $BV\ll1$.  If the field is tracking the matter density and $\mathcal D \approx 0$, then $V\sim \frac{1}{2}\dot\phi^2$ and both conditions hold simultaneously. The disformal factor evolves as
\begin{equation}\label{scalingD}
B\dot\phi^2 = 2^{\frac{\beta}{\alpha}}\left[\Omega_q  \rho M_p^{-4} \right]^{1-\frac{\beta}{\alpha}}
\propto a^{3(1-\frac{\beta}{\alpha})}\,,
\end{equation}
and will be a growing function of time only if $\beta/\alpha>1$, a condition necessary to push the field out of the attractor. 
The parameter $\phi_x$ produces a shift in the transition time and hence $\Omega_\phi$ is a monotonous growing function of it. In the following we will discuss the model in terms of $\alpha$, $\beta/\alpha$ and $\Omega_\phi$.


\subsubsection{Accelerated expansion}\label{late times}

When the disformal factor reaches $\mathcal O(1)$ values, the lapse function reduces significantly and the field, as seen in the Einstein frame, is pushed towards a slow roll phase. This produces a negative equation of state responsible for the acceleration of the universe. Figure \ref{background-plots} displays the effects of the transition in $w_\phi$ for different values of $\beta/\alpha$.  High values lead to a rapid slow down of the field and a more negative equation of state while low values are associated with gradual and longer transitions in which the field preserves a significant velocity at later times.  The reasons for this behavior and the equation of state will be analyzed in further detail below (\ref{transition-section}).

\FIGURE{
\includegraphics[width=0.49 \columnwidth]{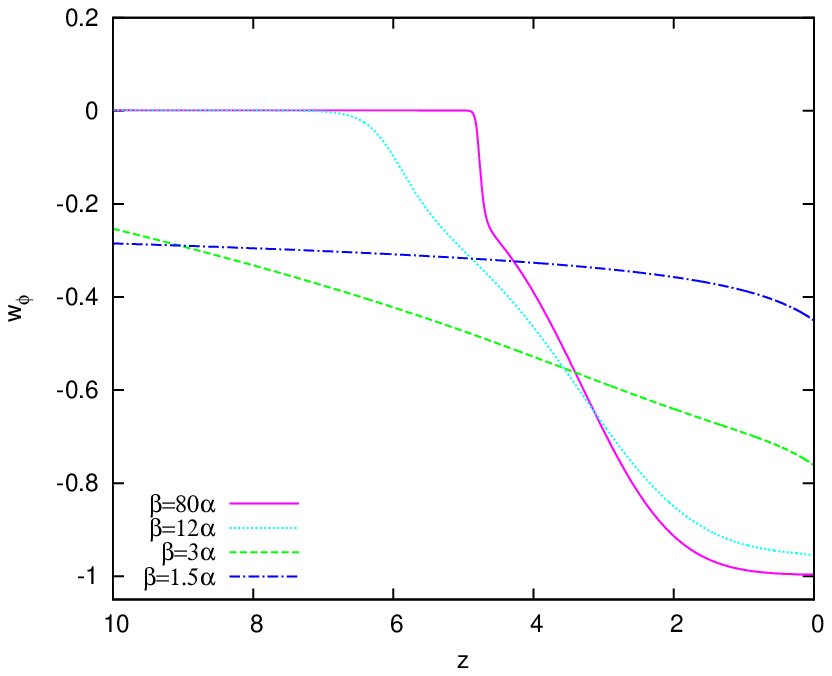}
\includegraphics[width=0.49 \columnwidth]{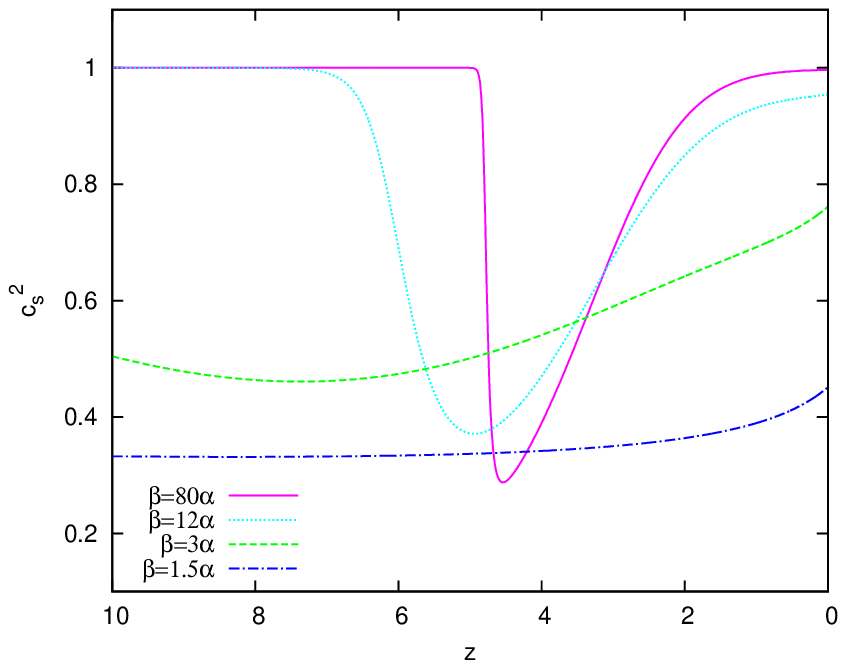}
\caption{Disformal quintessence dependence $\beta/\alpha$ as a function of redshift ($\alpha=10$ and $\Omega_q=0.7$).
\strong{Left:} Equation of state for the field. Higher values of $\beta/\alpha$ produce sharper transitions while low values lead to smoother ones.. \strong{Right:} Speed of sound squared (see section \ref{perturb-section}), equivalent to $\mathcal F/\mathcal M\approx 1-\mathcal D$.
\label{background-plots}} 
}

The disformal factor depends on both $\phi, \dot\phi$ and has a nontrivial behavior (See Figure \ref{disformal-stuff}). The dynamics are similar in all cases, but become more clear for higher values of $\beta/\alpha$. The transition is associated with a maximum and a rapid fall towards a certain value, followed by a slower reduction. The existence of a maximum follows from the role of $\mathcal D$ in the disformal metric, which implies that $\mathcal D<1$ at any time. Otherwise the metric would run into a singularity and many quantities would blow up, notoriously the energy density (\ref{rho}). A slow down of the field is therefore to be expected whenever $\mathcal D$ approaches unity.

\FIGURE{
\includegraphics[width=0.6 \columnwidth]{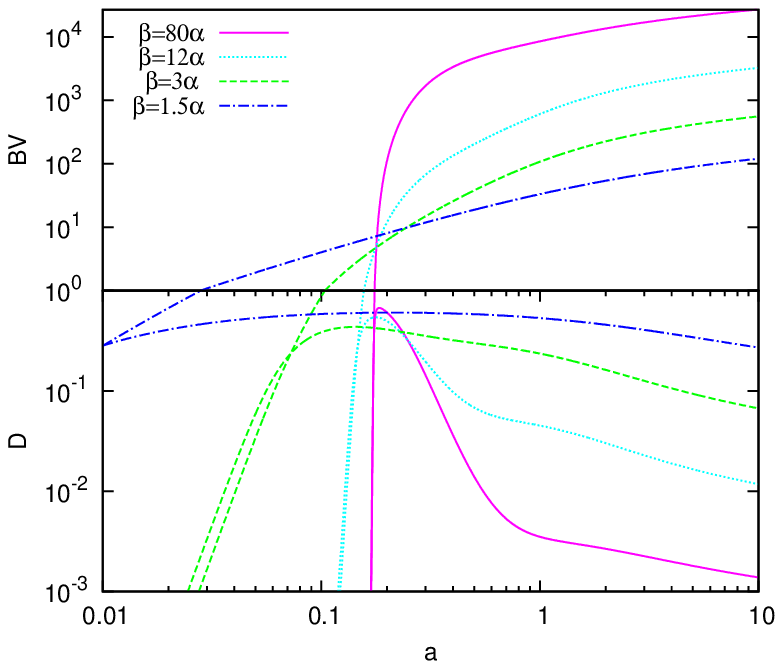}
\caption{Disformal functions $\mathcal D$, $BV$ for different $\beta/\alpha$ as a function of the scale factor ($\alpha=10$ and $\Omega_\phi=0.7$). Before the transition both lines evolve together. The disformal factor $\mathcal D < 1$ corresponds to the lower lines. The bounce avoids a metric singularity in (\ref{disf}). The disformal-potential term $BV$ corresponds to the lines that cross above one. This term dominates $\mathcal M$ after the transition. Note that the evolution has been extrapolated to $a=10$.
\label{disformal-stuff}} 
}

No similar bound can be put on the potential disformal factor, which grows exponentially with $\phi$ and can overcome the small velocity and reach large values after the freeze out. It can be seen how the change of slope in $BV$ roughly corresponds to the transition (i.e. the slow down of the field).
This factor is the dominant contribution to $\mathcal M$ and will produce a large inertia for the field. Despite the disformal factor is restored after the transition, the dynamics of the field might be strongly suppressed by the large value of $BV$. 

\subsubsection{Disformal transition in detail} \label{transition-section}

If $\beta \ll \alpha$, the disformal factor is well below one except during some interval around freeze out. This observation suggest an expansion to first order in $\mathcal D$ to gain some intuition about the model. Let us regard the terms in (\ref{equation}) in the light of this expansion. For the friction term  we get a simple correction with respect to the usual quintessence case ($\mathcal F/ \mathcal M=1$)
\begin{equation}
\frac{\mathcal F}{\mathcal M}=1-\mathcal D +\mathcal O(\mathcal D^2)\,.
\end{equation}

The expansion of $\mathcal P/\mathcal M$ taking into account the exponential dependence on the field yields
\begin{eqnarray}
\frac{\mathcal P}{\mathcal M} = -\frac{\alpha}{M_p}\frac{V}{1+BV}\left[1 + 3  \mathcal D  \frac{1-2BV}{2(1+BV)} \right] 
+\frac{\beta}{M_p}\mathcal D \frac{1}{2} \frac{\frac{3}{2}\dot \phi ^2 + V}{1+BV} +\mathcal O(\mathcal D^2)  \,. 
\label{PoM}
\end{eqnarray}
The first term reduces to the usual $V_{,\phi}$ at early times, but it will receive an additional suppression by the $BV$ factor in the denominator growing large at later times (The expression in brackets remains of order 1). The second term represents the purely disformal contribution to the potential and is equally suppressed by $BV$ at late times. Note that it has the proper sign to cause a deceleration of the scalar and is suppressed by the disformal factor prior to the transition.

Neglecting fine details from (\ref{PoM}), the force acting on the field will flip sign when
\begin{equation}\label{transition}
 \mathcal D \sim \frac{\alpha}{\beta}\,.
\end{equation}
This force associated with the transition can grow much larger than $V_{,\phi}$ if $\beta \gg \alpha$, which is the ultimate cause for the 
transition to begin so rapidly. As the field slows down, the disformal factor $\propto \dot\phi^2$ and the restoring force are reduced. 
Eventually, the usual term will come to dominate the force again when (\ref{transition}) is fulfilled again. We can see in the plots that the disformal factor stabilizes approximately around this value in Figure \ref{disformal-stuff}.

As the transition happens the field increases its value and the $BV$ factor in the denominator strongly suppresses the force term acting on the field. The later and softer slow down would then be driven by the friction term, which does not suffer this or other suppression.

\subsubsection{Equation of state} \label{eqofstate-section}

\FIGURE{
 \includegraphics[width=.6 \columnwidth]{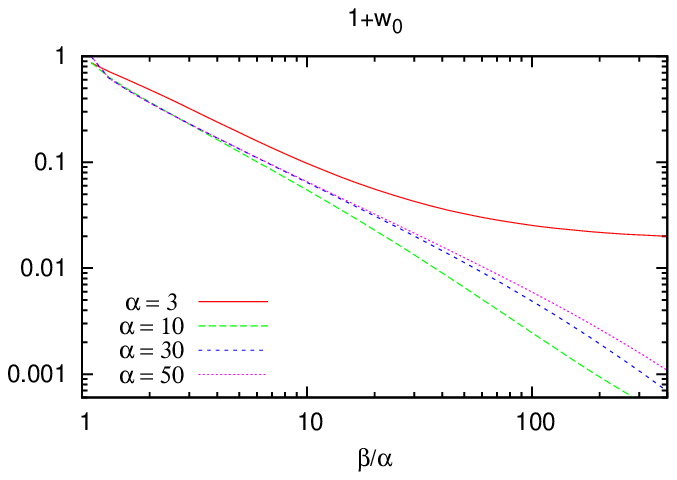}
\caption{Field equation of state at present time as a function of $\beta/\alpha$ for different values of $\alpha$. There is a clear trend towards $w_\phi-1+\alpha/\beta$ unless $\alpha$ or $\beta/\alpha$ are very low. For the red curve the amount of early dark energy is very large and the disformal transition is not finished at present time, hence the deviation.
\label{w_0}}
}

We can compute the equation of state from equations (\ref{pressure}, \ref{rho}) to first order in $\dot\phi^2/V$ and $\mathcal D$ to be $w \approx \mathcal -1 + \mathcal D + \frac{1}{2}\dot\phi^2/V$. At late times, when the kinetic energy is negligible, the equation of state becomes
\begin{equation}\label{eqofstate}
w \approx -1 + \mathcal D \,,
\end{equation}
and we expect that $w_0 +1 \sim \alpha/\beta$ according to the discussion above. Figure \ref{w_0} confirms this trend for a wide range of parameters. For larger amounts of early dark energy the freeze out does not finish before the required $\Omega_\phi$ is reached, as it becomes clear for the curve with $\alpha = 3$. Similar considerations apply for long lasting freeze out (low values of $\beta/\alpha$).

The properties of the accelerated expansion are therefore linked to the quotient $\beta/\alpha$. Apparently the only role of $\alpha$ is to regulate the amount of early dark energy and therefore shift the transition time to compensate the time interval necessary to achieve $\Omega_\phi$ today.


\subsection{Perturbations} \label{perturb-section}

We considered the perturbations in the synchronous gauge \cite{Ma:1995ey}. The relevant equations are the same as in \cite{Hwang:2005hb} were $p$ is given by the Lagrange density (\ref{unbarred}). With the definitions from the previous section, we can write the energy density and velocity perturbation induced by the disformal field as
\begin{equation} \label{delRhox}
\delta \rho_\phi = {\mathcal L^{-5/2}} \left[\mathcal M  \dot \phi \dot{\delta\phi} + {\mathcal P} \delta\phi\right] \,,
\end{equation}
\begin{equation} \label{delTheta}
(\rho+p)\theta_\phi = - \frac{\mathcal F}{\mathcal L^{5/2}} \frac{k^2}{a}  \dot \phi  \delta\phi \,.
\end{equation}
The Klein-Gordon equation obeyed by such perturbations can be written as
\begin{align} \label{kg-chachi}
 \ddot{ \delta \phi} + \left[3 H +  \frac{\dot{\mathcal M}}{\mathcal M} \right] \dot{\delta\phi}  
+ \left[ \frac{\mathcal F}{ \mathcal M }k^2 + m_\phi^2 \right] \delta \phi
 +    \frac{1}{2} \frac{\mathcal F}{ \mathcal M }  \dot h_k \dot \phi  = 0 \,.
\end{align}
$h_k$ is the interaction with gravity introduced through the covariant derivative. The field mass $m^2_\phi$ generalizes the $V_{,\phi\phi}$ term and is given in terms of (background valuated) partial derivatives of (\ref{unbarred}) as
\begin{equation}
m^2_\phi= 
\frac{1}{\mathcal M}\left[ (\ddot \phi +3 H \dot \phi)p_{,X\phi} + \dot p_{,X\phi} \dot \phi  + p_{,\phi\phi}\right]  \,.
\end{equation}


\subsubsection{Tracking}

As in the homogeneous limit, the smallness of the disformal factor ensures no significant departure from the canonical quintessence case at early times.

The speed of sound characterizes the propagation of fluctuations within the field. For the purely disformal case it is given by \cite{Koivisto:2008ak}
\begin{equation}
 c_\phi^2 =\frac{\mathcal F}{\mathcal M} = \mathcal L \frac{1+BV\mathcal L - \mathcal D/2}{1+BV\mathcal L + \mathcal D/2}\,.
\end{equation}
The allowed values of $BV$ and $\mathcal D$ bound its value between $0$ and $1$ (See Figure \ref{background-plots}). It is very close to one before the transition, according to the canonical quintessence limit. Therefore, the growth of perturbations at the tracking stage will be strongly suppressed and by the time of the transition the  dark energy overdensity will be many orders of magnitude below the matter contrast.

As for usual quintessence, the finite amount of early dark energy will increase the expansion rate and damp the formation of structure in all epochs. This departure of matter domination is reflected in a slope in the Newtonian gravitational potentials and a slight enhancement of the first acoustic peaks on the CMB. \cite{ferreira,skordis}.
 
\FIGURE{
\includegraphics[width=.48 \columnwidth]{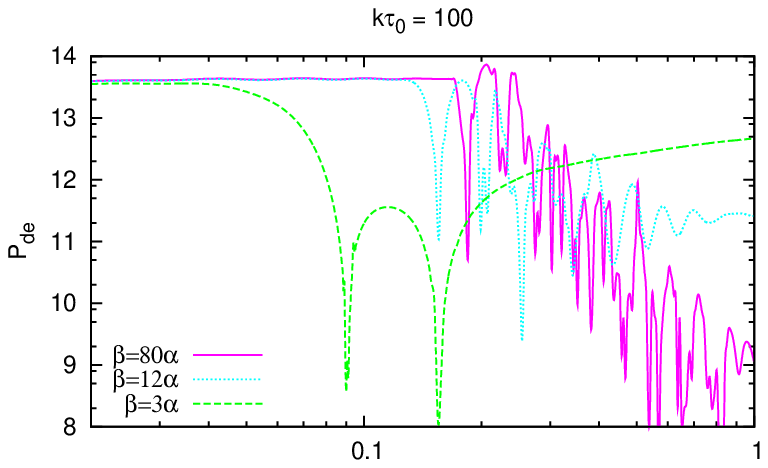}
\includegraphics[width=.48 \columnwidth]{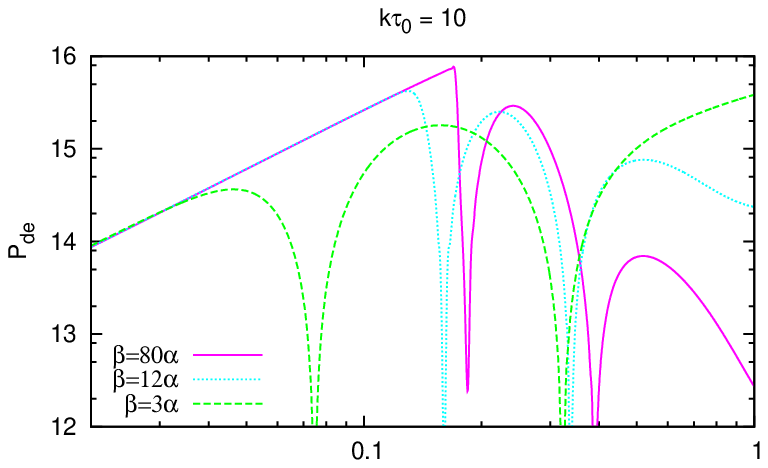} \\[.3cm]
\includegraphics[width=.48 \columnwidth]{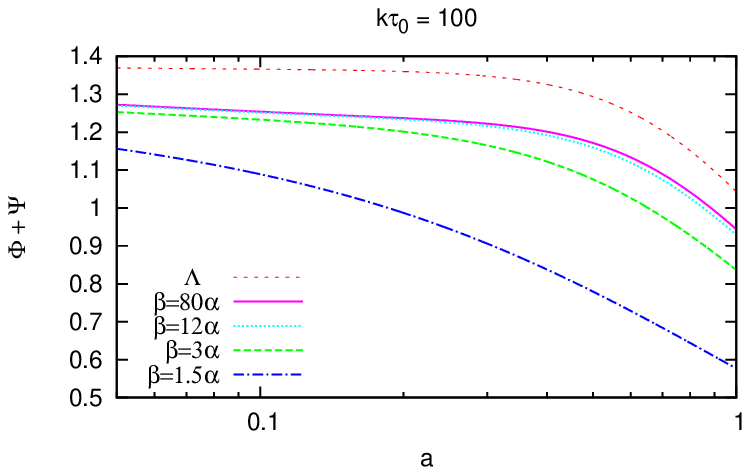}
\includegraphics[width=.48 \columnwidth]{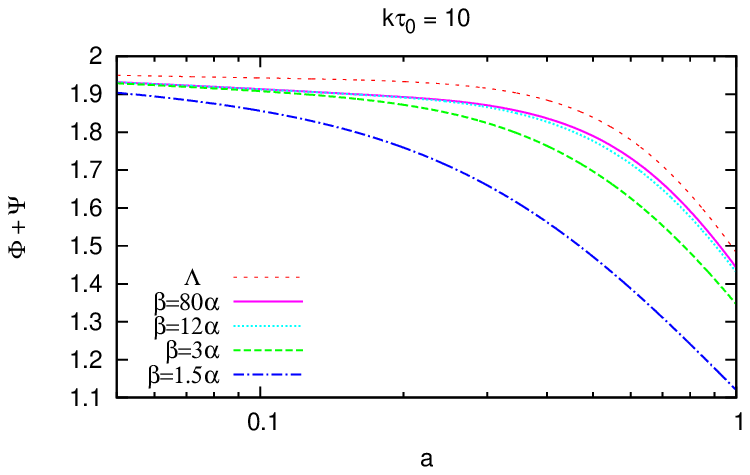}
\caption{Large scale perturbations for $\alpha=10$. \strong{Top:} Dark energy power spectrum (arbitrary units). Lower values of $\beta/\alpha$ allow more clustering due to different expansion rate and speed of sound. \strong{Bottom:} Gauge invariant potentials for the same $k-$modes. The main differences with $\Lambda$ are departure from matter domination due to early dark energy and the different times in which the acceleration epoch begins.} \label{Powergrowht}}

\subsubsection{Accelerated expansion}

The transition causes changes in the terms involved in (\ref{delRhox},\ref{delTheta},\ref{kg-chachi}) such as the sign flip and rapid growth in $\mathcal P$ associated with the freeze out. This variations are compensated by corresponding changes in $\delta\phi$, which reduces in magnitude and velocity to maintain the energy density at a similar level (see Figure \ref{Powergrowht}). The speed of sound drops below 1 for a certain time and then stabilizes with a value that depends on $\beta/\alpha$ (See Figure \ref{background-plots}). This would in principle enhance the clustering properties of the disformal field after the transition, but the inhomogeneities in the fluid can grow no faster than the matter ones, and since they are many orders of magnitude smaller at the beginning of the transition they do not play a relevant role. Besides, the transition is associated with the accelerated expansion and therefore the growth of structure is further suppressed. 

The main departures from \LCDM are caused by the different expansion rates in the accelerated phase and the different transition times. Lower values of $\beta/\alpha$ render the transition slower, but a longer expansion period is required to achieve the same density fraction. This both affects the Integrated Sachs Wolfe effect and shifts the angular scales of the CMB. Higher values mimic the $\Lambda$ behavior at late times, and will only reflect the departures from matter domination described above for small $\alpha$, i.e. significant early dark energy. 

\FIGURE{
\includegraphics[width=.6\columnwidth]{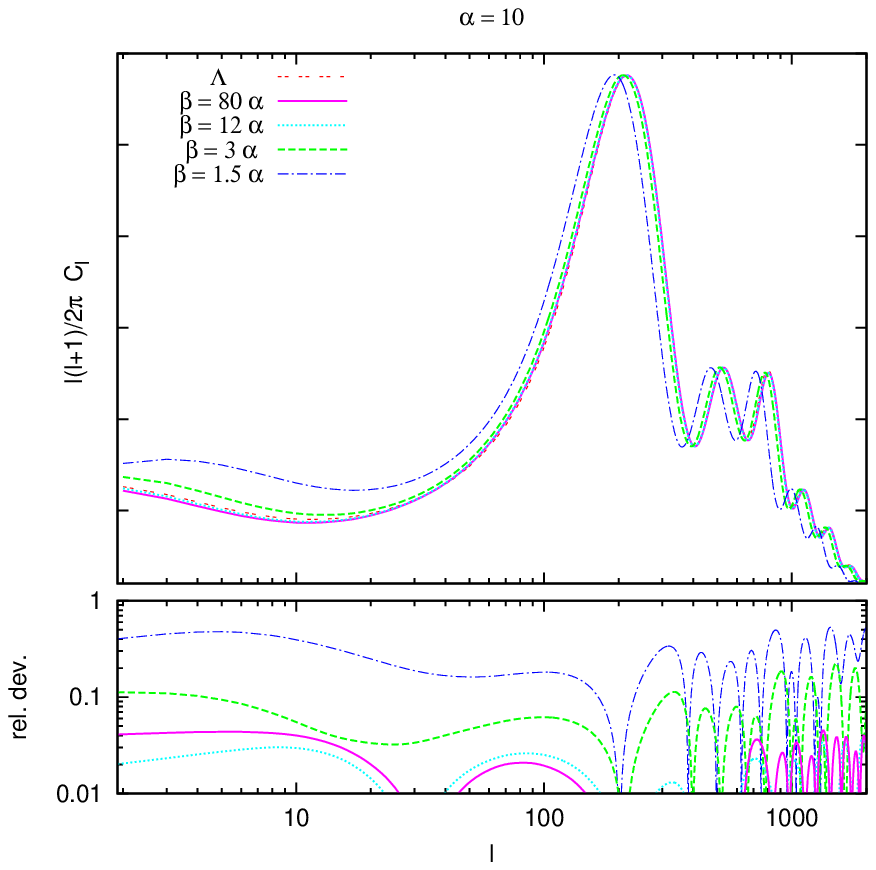}
\caption{CMB spectra for different values of $\beta/\alpha$ ($\alpha=10$) and relative deviation with respect to $\Lambda$. The initial amplitude of the perturbations has been chosen to correct the amount of early dark energy and fit the first peak. The departures from \LCDM are noticeable as a shift of the angular scale and the increase of the Sachs-Wolfe plateau, but is only significant for very low values of $\beta/\alpha$.} \label{CMB}
}

\FIGURE{
\includegraphics[width=.6\columnwidth]{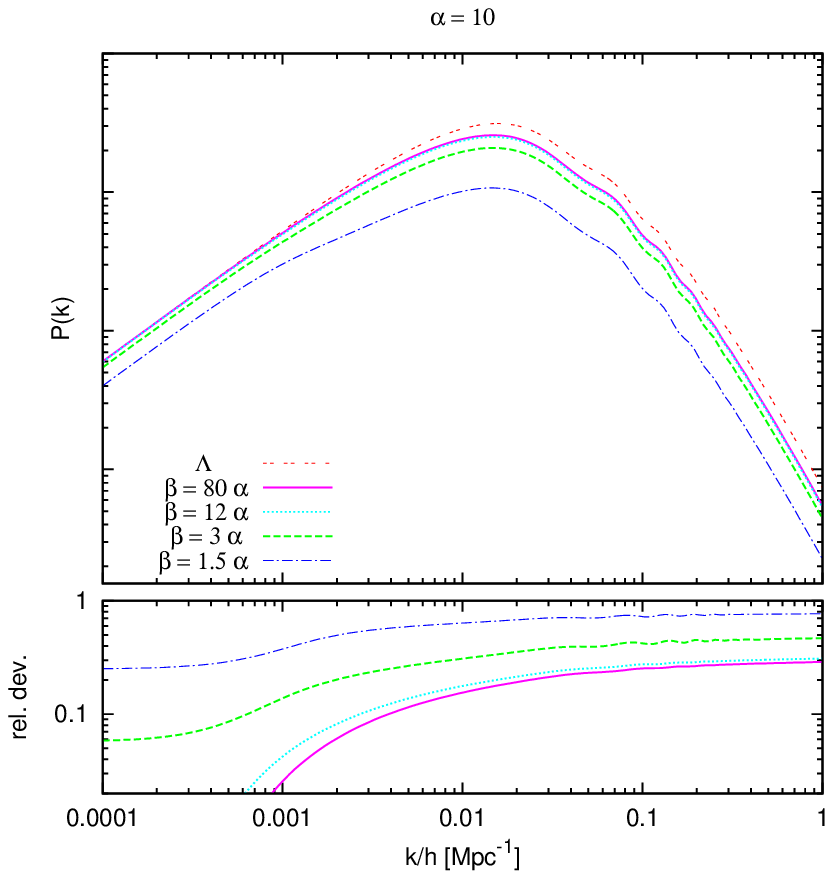}
\caption{CDM spectra for different values of $\beta/\alpha$ ($\alpha=10$) and relative deviation with respect to $\Lambda$. 
The spectra fail to converge to $\Lambda$ in the limit of high $\beta/\alpha$ due to the finite amount of early dark energy at $\alpha=10$, which damps cosmic growth. As the shape of the spectrum remains the same, it is difficult to obtain constraints from it due to the ignorance of the linear galaxy bias.} \label{CDM}
}

\subsection{Constraints} \label{constraints-section}

A Monte Carlo Markov Chain simulation was used to obtain constraints on the parameter space of the theory by comparing with WMAP 7 year data \cite{wmap}, supernovae from the Union dataset \cite{sn1}, SDSS DR7 baryon acoustic peak position \cite{bao7} and matter power from the SDSS luminous red galaxies sample \cite{lrg}. The uncertainty in the linear galaxy bias factor was treated by choosing the highest likelihood value in the range $(1,3)$. This choice accounts for the uncertainty associated while avoiding too large bias factors. Supernova data included systematic errors and possible spatial curvature was neglected. The upper bounds in the model parameter space $\alpha\in(3,20)$ and $\beta/\alpha\in(1.5,30)$ was chosen by sensitivity considerations, since very little departure from $\Lambda$ can be observed at these values. The lower bounds ensure the existence of matter attractor solution for the field ($\alpha>3$), disformal transition ($\beta>\alpha$) and avoid computational difficulties. The simulation was performed through a modification of the publicly available Boltzmann code CMBeasy \cite{cmbeasy,cmbeasymcmc}. Four chains were run adding up to 20870 accepted models. An additional MCMC using only background information (SNe, BAO and CMB distance priors as described in section 5 of \cite{wmap}) was run for comparison. Results are displayed in Table \ref{results} and Figure \ref{contours} (shaded regions and black lines for the full and restricted dataset respectively).

\FIGURE{
\includegraphics[width=.49\columnwidth]{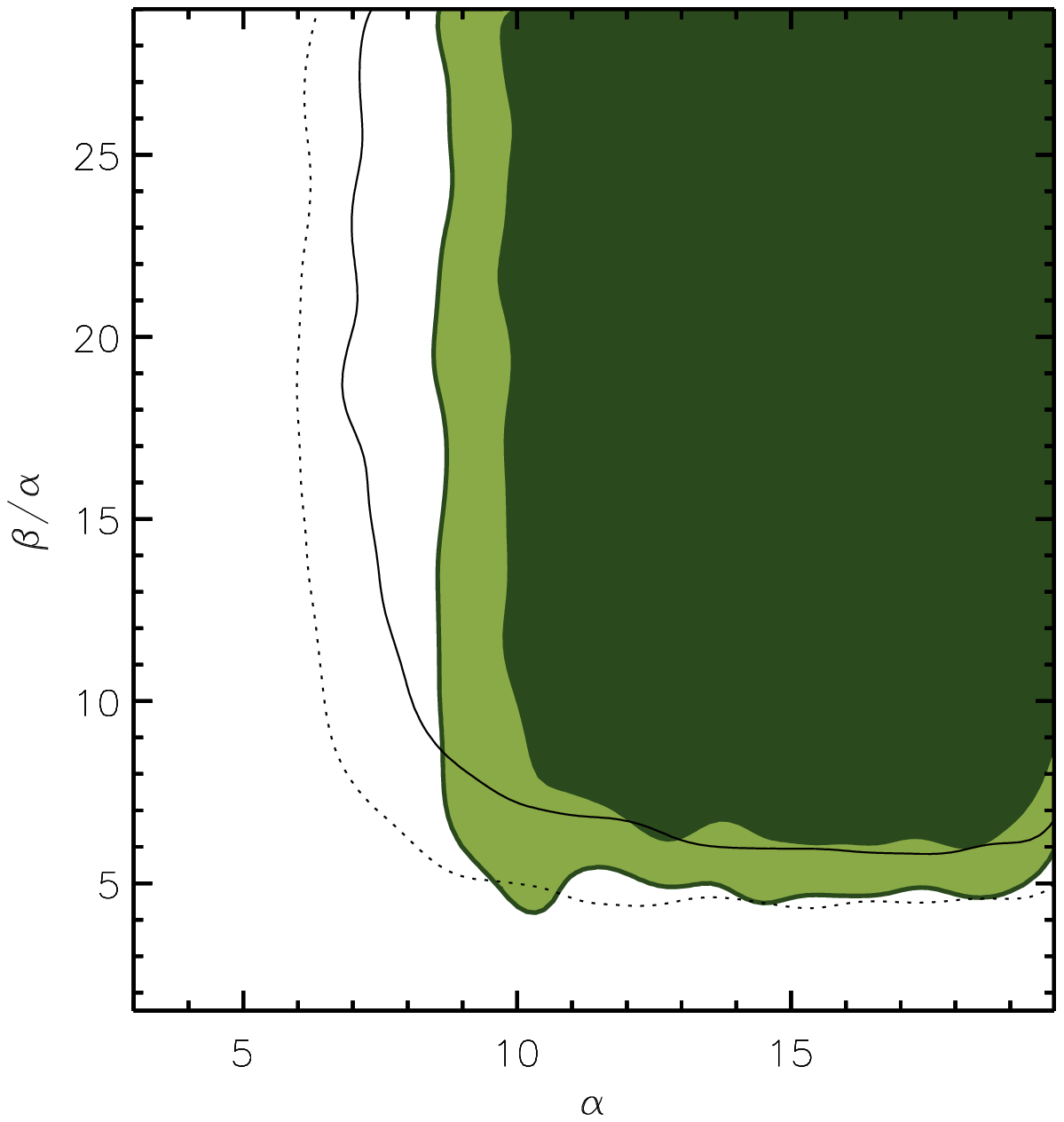} 
\includegraphics[width=.49\columnwidth]{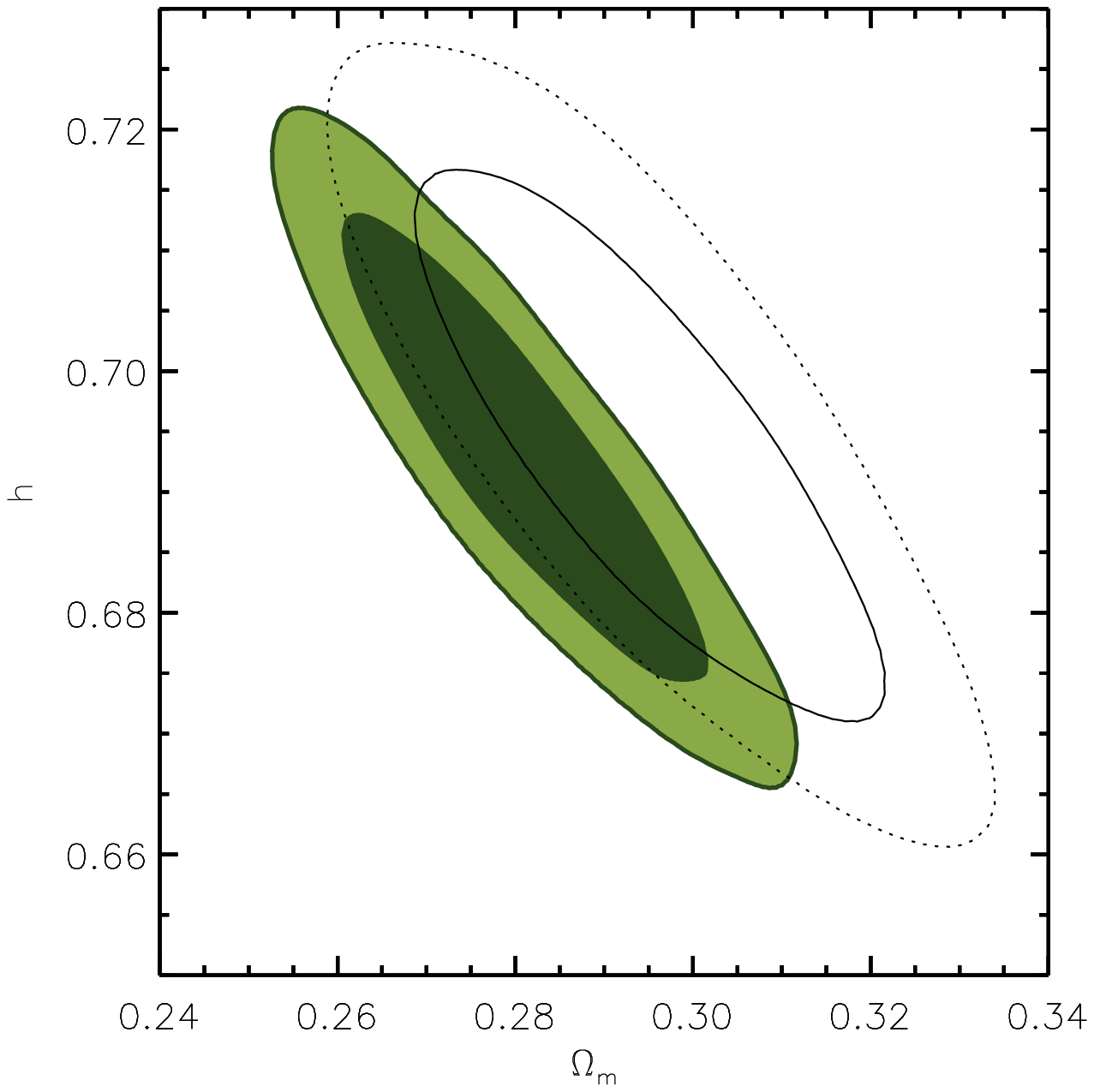} 
\caption{Likelihood Contours for disformal quintessence. Shaded regions correspond to the 1 and 2-$\sigma$ obtained by the full MCMC analysis while the black lines correspond to a restricted analysis using only information from the background expansion (SNe, BAO and CMB distance priors). The constraints in $\alpha$ and $\beta/\alpha$ correspond respectively to high fractions of early dark energy and slow disformal transitions towards accelerated expansion.} \label{contours}
}

\TABLE{
\begin{tabular}{c  c l  }
Parameter \ \ &  Maximum Likelihood \ \  & 68\% C.L.  \\
\hline 
$100\Omega_b h^2 $ & $ 2.23 $ &  $\pm 0.05$  \\[3pt]
$\Omega_m h^2 $ & $ 0.135 $ &  $\pm 0.004 $  \\[3pt]
$h$ & $ 0.690 $ & $ \pm {0.014}$  \\[3pt]
\hline
$\alpha$ & $ 15.9 $ & $\gtrsim 8.5 $ \\ [3pt]
$\beta /\alpha$ & $ 17.2 $ &  $\gtrsim 4.9 $ \\[3pt]
\hline
$\tau $ & $ 0.086 $ &  $\pm 0.015 $  \\[3pt]
$n_s $ & $ 0.971 $ &  $ \pm 0.014 $  \\[3pt]
Amp. & $ 2.91 $ & $ \pm 0.015$    \\[3pt]
\hline
\end{tabular}
\caption{Results from the Monte Carlo Markov Chain analysis. All the usual cosmological parameters lie within one sigma from the WMAP 7 constraints presented in \cite{wmap}. Amplitude is given by $\ln (10^{10} A_s) - 2\tau$. \label{results}}
}

Observations are compatible with a large patch of parameter space in the $\alpha,\beta/\alpha$ plane and confirm the preference towards higher values of both parameters where the $\Lambda$ limit can be achieved. Constraints are tighter on $\alpha$ due to the cumulative effect that early dark energy exerts in the growth of perturbations, as can be inferred by a comparison between the runs using the complete and the restricted sets of observations (top plot in Figure \ref{contours}). This difference is likely to arise when the allowed bias factor fails to compensate the effect of early dark energy to produce a good fit to LRG, as well as due to the effect in the third and further acoustic peaks (Figures \ref{CDM} and \ref{CMB}). The constraints on the properties of accelerated expansion (i.e. on $\beta/\alpha$) do not improve significantly in the full analysis because the relevant features affect mostly distance information, encoded in SNe, BAO and CMB angular shift measurements. Other features such as the integrated Sachs-Wolfe effect are significant only for very low values of the parameters and do not introduce relevant differences in the region of interest (Figure \ref{CMB}).
There is a slight mismatch between both runs in the determination of $\Omega_m$ (second plot of Figure \ref{contours}). It seems unlikely that the discrepancy is caused by passing from the full WMAP7 spectra to the homogeneous distance priors $z_*,l_A,\mathcal R$. Though less restrictive, distance priors should essentially agree with the complete data. The shift would be due to the additional information of the matter power spectrum from LRG. The lower value of $\rho_m$ would raise the matter power whenever the allowed linear bias is not large enough, since $P_k \propto \rho_m ^{-2}$.


\section{Conclusions}
\label{conclusions-section}

Disformal generalizations of the conformal transformation have found several applications in cosmology, particularly in the 
frameworks of gravitational alternatives to dark matter and varying light speed alternatives to inflation. In this paper we have shown that 
the disformal relation can be also very prolific in generating alternative explanations for the cosmic acceleration, suggesting new 
links between models worth further explorations. In the future, it would be interesting to study the disformal relations which may exist 
between more general field theories. Already the details of motion of point particles in a disformal metric are, to our knowledge, not very 
well understood. Whilst the nature of dark energy is undisclosed and it is not known to which metric it is coupled, there is in fact 
no fundamental principle establishing the precise relation between the gravitational metric and the physical metric in which ordinary 
matter lives \cite{bi,bi2,bi3,teves,aether}. It is thus useful to investigate the observational signatures and put experimental bounds on it from e.g. classical tests of the equivalence principle.

Here we considered the canonical scalar field action where metric is replaced by the disformal one, i.e. made the substitution  
$g_{\mu\nu}\to\bar g_{\mu\nu}$ in the Lagrangian. At various limits we can then obtain Chaplygin gas and other models considered in the 
literature. Disformal quintessence arises in a very simple way as an application of this substitution, which results effectively in a 
non-standard self-interaction of the scalar field. This self-interaction causes the field to accelerate the universe given exponential functions of the field and under the condition $\alpha/\beta<1$ in (\ref{B},\ref{V}). The acceleration can begin near the present epoch if both $\alpha$ and $\beta$ are of order one and $\phi_x\sim M_p$, meaning that {\it all} the parameters appearing in the Lagrangian can be set to the Planck scale. For the magnitude of the potential and the disformal factor this follows from the special shift symmetry property of the exponential form. Moreover, as well known, in 
cosmology the exponential potential has the special scaling property, so practically no tuning whatsoever of initial conditions for the 
field is required. 
These observations motivated us to consider the phenomenology of the model in detail and to confront it with the latest available 
data.

We considered evolution both at the level of background and of linear perturbations. The dependence of the background dynamics on 
model parameters allows us to put constraints on them. These become more stringent if we take into account the evolution of linear structures, 
though the effect of the perturbations of the field to them is more subtle.   
A Monte Carlo Markov Chain simulation was used to obtain constraints on the parameter space of the theory by comparing with WMAP 7 year data, 
SNe from the Union dataset, baryon acoustic peak position and luminous red galaxies power spectrum from SDSS. Six cosmological 
parameters were allowed to vary together with $\alpha$ and $\beta$. 

It was then found that small values of $\alpha$ can be ruled out because of the
effect of early dark energy that implies. On the other hand, small values of $\beta/\alpha$ result in slow transition to acceleration, 
which is also disfavored by the data. However, when both $\alpha$ and $\beta/\alpha$ are sufficiently large (the precise limits shown 
in Figure \ref{contours}), it becomes very difficult to distinguish the model from $\Lambda$CDM and thus this parameter region 
remains compatible with the present data. These constraints might be improved by considering nonlinear scales and gaining a deeper understanding of the effects on large scale structure.

Further research might focus on the role of the disformal relation in dark energy scenarios by the introduction of couplings to other forms of matter and gravity constructed by following the disformal prescription $g_{\mu\nu}\to\bar g_{\mu\nu}$ as well the relation of disformal quintessence with other forms of dark energy and non-minimal derivative coupled version of scalar tensor theories.


\acknowledgments

We would like to thank G. Robbers for his advice while modifying CMBeasy.
TK is supported by the Finnish academy and the Yggdrasil grant of the Norwegian Research Council. MZ is supported by MICINN (Spain) through the project AYA2006-05369 and the grant BES-2008-009090. TK and MZ are grateful for the Oslo Institute of Theoretical Astrophysics, and MZ to the Heidelberg Institute of Theoretical Physics for hospitality during the completion of this work. DFM is supported by the Norwegian Research Council FRINAT grant $197251/V30$.

\bibliography{refs}

\end{document}